\documentclass[conference]{IEEEtran}
\usepackage[printonlyused]{acronym}
\usepackage{adjustbox}
\usepackage[style=ieee,backend=bibtex]{biblatex}
\bibliography{biblio}
\include{macros}
\usepackage{array}

\usepackage[font=small]{caption}
\usepackage{subfigure}
\usepackage{bm}
\usepackage{amsmath}
\usepackage{breqn}
\usepackage{mathtools}

\DeclarePairedDelimiter\floor{\lfloor}{\rfloor}

\begin{document}

\title{Intelligent Reflective Surface vs. Mobile Relay-supported NLoS Avoidance in Indoor mmWave Networks}
\vspace{-3mm}
\author{
\IEEEauthorblockN{Maria Bustamante Madrid, Jeroen Famaey, Filip Lemic}
\IEEEauthorblockA{Internet Technology and Data Science Lab (IDLab), Universiteit Antwerpen - imec, Belgium \\
Email: \{maria.bustamantemadrid, filip.lemic, jeroen.famaey\}@uantwerpen.be}
\vspace{-6mm}
}

\maketitle

\begin{abstract}
The 6$^{\bm{th}}$ generation of wireless communication (6G) is envisioned to give rise to various technologies for improving the end-to-end communication performance, where the communication is envisioned to utilize wireless signals in the millimeter wave (mmWave) frequencies and above. 
Among others, these technologies comprise \acp{IRS} and \acp{MR}, whose envisaged roles include mitigating the negative effects of \ac{NLoS} connectivity, in particular at mmWave and higher frequencies. 
The core idea behind these technologies is to use cooperative networking where the source sends a signal to a repeater, in this case the IRS or the MR, which is upon reception forwarded to the destination.
When comparing the two technologies, it is important to realize that the IRSs are primarily envisioned to be static entities attached to various objects in the environment such as walls and furniture.
In contrast, the MRs will feature a higher degree of freedom, as they will be able to position themselves seamlessly in the environment. 
Based on the above assumptions, we derive an approach for determining the optimal position of the IRS and MR in indoor environments, i.e., the one that maximizes the end-to-end link quality between the source and the destination. 
We follow by capturing the communication quality indicators for both IRS- and MR-supported NLoS avoidance in indoor mmWave communication in a number of scenarios.  
Our results show that, from the end-to-end link quality perspective, the MRs generally outperform the IRSs, suggesting their utilization potential for throughput-optimized NLoS avoidance scenarios.
\end{abstract}



\acrodef{FSPL}{Free-Space Path Loss}
\acrodef{LoS}{Line-of-Sight}
\acrodef{NLoS}{Non-Line-of-Sight}
\acrodef{MR}{Mobile Relay}
\acrodef{UAV}{Unmanned Aerial Vehicle}
\acrodef{IRS}{Intelligent Reflective Surface}
\acrodef{RIS}{Reconfigurable Intelligent Surface}
\acrodef{SDM}{Software-Defined Metasurface}
\acrodef{SNR}{Signal-to-Noise Ratio}
\acrodef{AD/DA}{Analog-to-Digital/Digital-to-Analog}
\acrodef{LTI}{Linear Time-Invariant}
\acrodef{DF}{Decode-and-Forward}
\acrodef{AF}{Amplify-and-Forward}
\vspace{-1mm}

\section{Introduction}

We are witnessing the emergence of 5G, while simultaneously different 6G technologies are being explored for even faster and more reliable communications~\cite{di2020reconfigurable}. 
The demand for ever increasing data throughput and reliability can be motivated by the expectation of a monthly increase in the global mobile data traffic by 77~exabytes in 2022~\cite{di2020reconfigurable}. 
To support these demands, the migration to higher frequencies such as the millimeter wave (mmWave, 30-100~GHz) and above has been considered~\cite{di2020reconfigurable,struye2021millimeter}.
This is because of the available bandwidth in mmWave and above frequencies that offers the possibility of reaching the high data rates demanded by future 6G wireless communication systems. 
However, compared to the currently utilized sub-6~GHz, mmWave frequencies have higher \ac{FSPL} in \ac{LoS} propagation. 
More importantly, in \ac{NLoS} propagation environments such as indoors, the mmWave attenuation is higher than for the sub-6~GHz-operating systems due to the high sensitivity to blockage from obstacles that may exist in a deployment environment. 
This can significantly degrade the communication performance, introducing unstable and unreliable connectivity~\cite{struye2021millimeter,di2020reconfigurable}.

For this reason, it is necessary to create new propagation routes, so that the communication between the source and the destination is stable and reliable. 
One possibility suggested by the scientific community is to use a \acf{MR}~\cite{chen2013mobile,raghothaman2011system,lemic2017location}, as the \ac{MR} can transform an \ac{NLoS} communication link into two (or more) \ac{LoS} links. 
The \ac{MR} is a technology that needs to be equipped with an energy supply and front-end circuitry for the reception, processing, and retransmission of electromagnetic waves. 
This results in an increase in the consumption of energy and capital investment needed for deploying such networks~\cite{di2020reconfigurable}.
In contrast, they have the advantage of having a high degree of freedom in terms of their positioning in a deployment environment. 
In addition, they can be supported by the emerging paradigm of \acp{UAV}-supported wireless infrastructures, e.g.,~\cite{mendes2021small}. 

Another option is to utilize \acfp{IRS}, also often called \acp{RIS}~\cite{bjornson2021reconfigurable} and \acp{SDM}~\cite{liaskos2018new}. 
Explained in a rather general way, the \ac{IRS} is a surface formed by reflective passive elements, where each element is able to control the phase of the incident electromagnetic waves. 
By implementing these devices, the propagation of electromagnetic waves between the source and the destination can be configured flexibly, which provides a solution for signal deterioration.
The utilization of \acp{IRS} would imply a cost reduction with respect to the previously mentioned \acp{MR}, both in terms of capital investment and energy consumption. 
That is because the IRS elements simply passively reflect the signals without a need for radio frequency transmission~\cite{wu2021intelligent}.

Therefore, the main difference between these two technologies is that the \ac{MR} actively processes the received signal before retransmitting an amplified signal, while the \ac{IRS} passively reflects the signal without amplification~\cite{di2020reconfigurable}.
Both technologies are currently witnessing a rapid uptake in NLoS-avoidance scenarios, in particular for the mmWave communication targeted in this article.
For example, Bj{\"o}rnson~\emph{et al.}~\cite{bjornson2019intelligent} compared the performance of IRSs and \textit{static} \ac{DF} relays, concluding that very high rates and/or large IRSs will be needed for outperforming DF relaying, even in case of a static relay with comparatively lower degree of freedom in terms of its optimal positioning than the MR assumed in this work.
Moreover, \cite{stratidakis2021analytical} proposes a framework for optimal positioning of IRSs in indoor environments, demonstrating that such positioning can significantly improve the communication quality compared to its sub-optimal alternatives.   

Taking inspiration from the works above, in this paper we attempt to make a fair comparison of the performance of the \ac{MR} and the \ac{IRS}-supported NLoS avoidance of mmWave communication in indoor environments. 
We do that by modeling different NLoS avoidance strategies through both IRS- and MR-supported communication under the assumption that the IRS and MR are positioned in a way that provides a quasi-optimal SNR at the communication link between the source and the destination.
We support the quasi-optimal positioning of the IRS and MR through an exhaustive search at different potential locations of the IRS and MR under the constraints stemming from a given technology.
In other words, in the IRS-supported NLoS avoidance, we assume the IRS can be positioned anywhere on the outer walls of the considered environments, while the MR can be positioned at any location in the environment of interest.
Our results demonstrate that the \ac{MR} generally outperforms the \ac{IRS} in terms of the communication quality characterized by the average \ac{SNR} of the communication links between the source and the destination.


\section{System Overview}

\subsection{\acl{IRS}}
\begin{figure}[!t]
\centering
\includegraphics[width=.8\columnwidth]{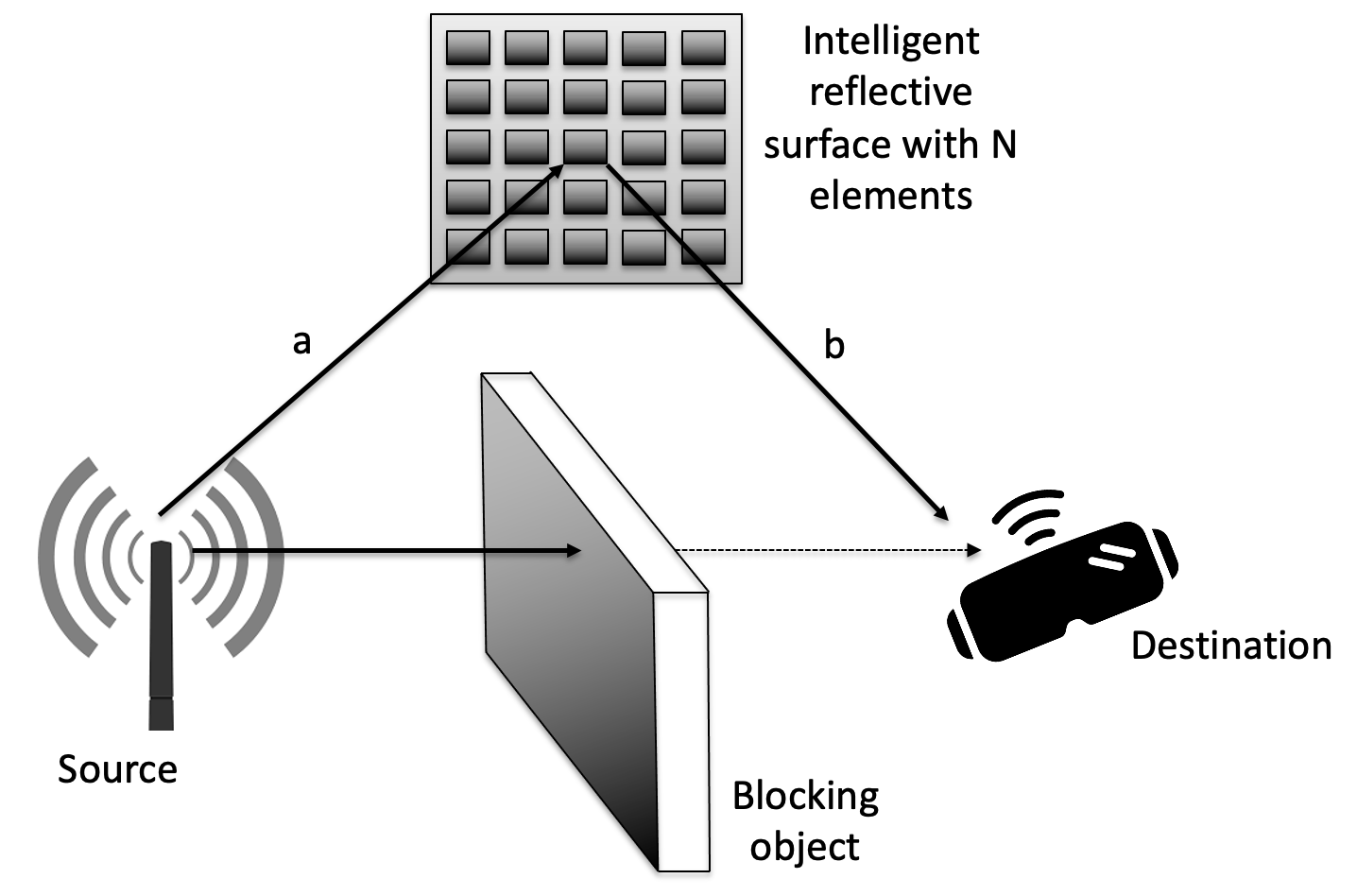}
\vspace{-1mm}
\caption{IRS-supported NLoS avoidance}
\label{fig:setup_irs}
\vspace{-3mm}
\end{figure}

The \ac{IRS} is capable of controlling the propagation of the waves that collide with it. 
The implementation of the IRSs can be done in two ways, i.e., by implementing large arrays of antennas whose inter-distance is in the order of the wavelength of the signals controlled, or by using metamaterial elements whose sizes and inter-distances are much smaller than the signal wavelength. 
In this work, we rely on the \ac{IRS} implementation based on metasurfaces~\cite{di2020reconfigurable}.
In general terms, the \ac{IRS} made of meta-surfaces is a matrix of a large number of passive reflective elements called meta-atoms as illustrated in Figure~\ref{fig:setup_irs}. 
They are smaller than the wavelength of the operating frequency of interest. 
Each element is capable of its individual signal response, being able to control the amplitude of reflection and the phase shift~\cite{bjornson2021reconfigurable,wu2021intelligent,wu2019towards}.

The \acp{IRS} provide characteristic advantages. 
Specifically, the \acp{IRS} can be implemented seamlessly. 
For example, the \acp{IRS} can be used to coat objects such as building facades, roofs, furniture, clothing, etc~\cite{lemic2021localization}. 
In addition, \acp{IRS} are environmentally friendly since they are almost passive and consume substantially less energy compared to conventional wireless systems. 
Additionally, the IRSs do not require \ac{AD/DA} converters or power amplifiers, hence they are cost effective~\cite{yuan2021reconfigurable}.

We consider the setup in Figure~\ref{fig:setup_irs}, focusing on the channel controllable by the \ac{IRS} and ignoring the direct \ac{LoS} between the source and the destination. 
We assume lossless beam-forming operation~\cite{rehman2021joint} for IRS-supported NLoS avoidance. 
A square-shaped \ac{IRS} has been selected for simplicity, while the generalization to rectangular shapes can be done by iterating through the antenna elements along X and Y-axes.  
For each \ac{IRS} element, we assume that the transmitted signal $x_{pb}(t)$ will be transmitted to the \ac{IRS} through a \ac{LTI} channel represented by an impulse response $a_{n,pb}(t)$, where $n = 1 ,. . . , N$ is the n$^{th}$ antenna element of a square-shaped \ac{IRS}.
When the signal reaches the element $n$, it is reflected and, therefore, irradiated again. 
The signal re-radiated by each element that forms the \ac{IRS} is radiated to the destination through another \ac{LTI} channel with an impulse response $b_{n,pb}(t)$. 
The joint impulse response is the convolution of the impulse responses for each IRS element $n$~\cite{bjornson2021reconfigurable}.
Considering the above, the \ac{SNR} at the IRS-supported communication link between the source and the destination can be written as~\cite{bjornson2020power}:

\begin{equation}
\label{eq:snr_irs}
SNR_{IRS} = \frac{P_{Tx}}{\sigma^2}\left(\sum_{n=1}^{N}{|a_n||b_n|}\right)^2,
\end{equation}

where $P_{Tx}$, $B$, and $n \sim N_c(0,\sigma^2)$ are the transmit power, utilized bandwidth, and the power of the additive white Gaussian noise at the destination, respectively.
Equation~\ref{eq:snr_irs} suggests that the SNR grows with $N^2$ when using the IRS that is optimally configured~\cite{bjornson2020power}. The values of $|a_n|$ and $|b_n|$ can be derived from the following (cf.,~\cite{bjornson2020power}):

\begin{equation}
|a_n|^2 = \xi_{(p_s),(p_n),\sqrt{A}},
\end{equation}

\begin{equation}
|b_n|^2 = \xi_{(p_d),(p_n),\sqrt{A}},
\end{equation}

where $\xi$ is the free-space channel gain, $p_s$, $p_r$, and $p_n$ are the positions of respectively the source, destination, and the IRS located at the $XY$-plane and centered at $p_n = (x_n, y_n, 0)$.

Based on~\cite{bjornson2020power}, we consider a lossless isotropic antenna located at $p_t = (xt , yt , d)$ that transmits a signal polarized in $Y$ direction when traveling in $Z$ direction. 
The receiver is located in the $XY$-plane centered at $p_n = (x_n , y_n , 0)$ and has the area $a\times a$. 
The free-space channel gain follows as:

\footnotesize
\begin{dmath}
\xi_{(p_t),(p_n),a} = \frac{1}{4\pi}\sum_{x \in X_{t,n}}{\sum_{y \in Y_{t,n}}}\left(\frac{xy/d^2}{3(y^2/d^2+1)\sqrt{x^2/d^2+y^2/d^2+1}} +\frac{2}{3}tan^{-1}{\frac{xy/d^2}{\sqrt{x^2/d^2+y^2/d^2+1}}}\right),
\end{dmath}
\normalsize

where $X_{t,n}=(a/2+x_n-x_t, a/2-x_n +x_t)$ and $Y_{t,n} = (a/2 + y_n - y_t, a/2 - y_n + y_t)$~\cite{bjornson2020power}. 
This equation is only true when the antenna area is significantly smaller compared to the wavelength, which is the case for the assumed \ac{IRS}.

To calculate $x_n$ and $y_n$, we analyze the position of the \ac{IRS} with respect to the source and the destination. 
Without any loss of generality, we assume the source is at the distance $d$ at angle $\eta$, while the destination is at distance $\delta$ in angle $\omega$ with respect to the IRS.
The position of the source is, therefore, equal to $p_s = (d\sin{\eta}, 0, d\cos(\eta))$ and that of the destination $p_d = (\delta\sin(\omega), 0 , \delta\cos(\omega))$. 
The location of each antenna element, expressed by its coordinates $x_n$ and $y_n$ for $n = 1, ..., N$, is then given as:

\begin{equation}
x_n = -\frac{(\sqrt{N}-1)\sqrt{A}}{2}+\sqrt{A}\mod{(n-1,\sqrt{N})},
\end{equation}

\begin{equation}
y_n = \frac{(\sqrt{N}-1)\sqrt{A}}{2}-\sqrt{A} \floor*{\frac{n-1}{\sqrt{N}}},
\end{equation}

where $N$ is the total number of elements in the \ac{IRS} and $A$ the area of each element.

\subsection{\acl{MR}}
The \ac{MR}, like the \ac{IRS}, is in charge of maintaining the \ac{LoS} path from the source to the destination. 
The most common retransmission strategies include \acf{DF}, and \ac{AF}. 
The \ac{DF} \ac{MR} decodes and remodulates the received signal, and retransmits it towards the destination, while the \ac{AF} \ac{MR} solely amplifies the received signal in a way that is transparent to the destination, and retransmits it without decoding it~\cite{levin2012amplify,bjornson2020reconfigurable}.
The \acp{MR} operate in half duplex, which means that the reception and transmission are separated in time. 
Currently, full duplex \ac{MR} capable of receiving and transmitting simultaneously are emerging~\cite{bjornson2020reconfigurable,nguyen2019closed}. 
In this article, we are nonetheless going to focus on the use of \ac{DF} \ac{MR} that operates in the half-duplex mode.

\begin{figure}[!t]
\centering
\includegraphics[width=.8\columnwidth]{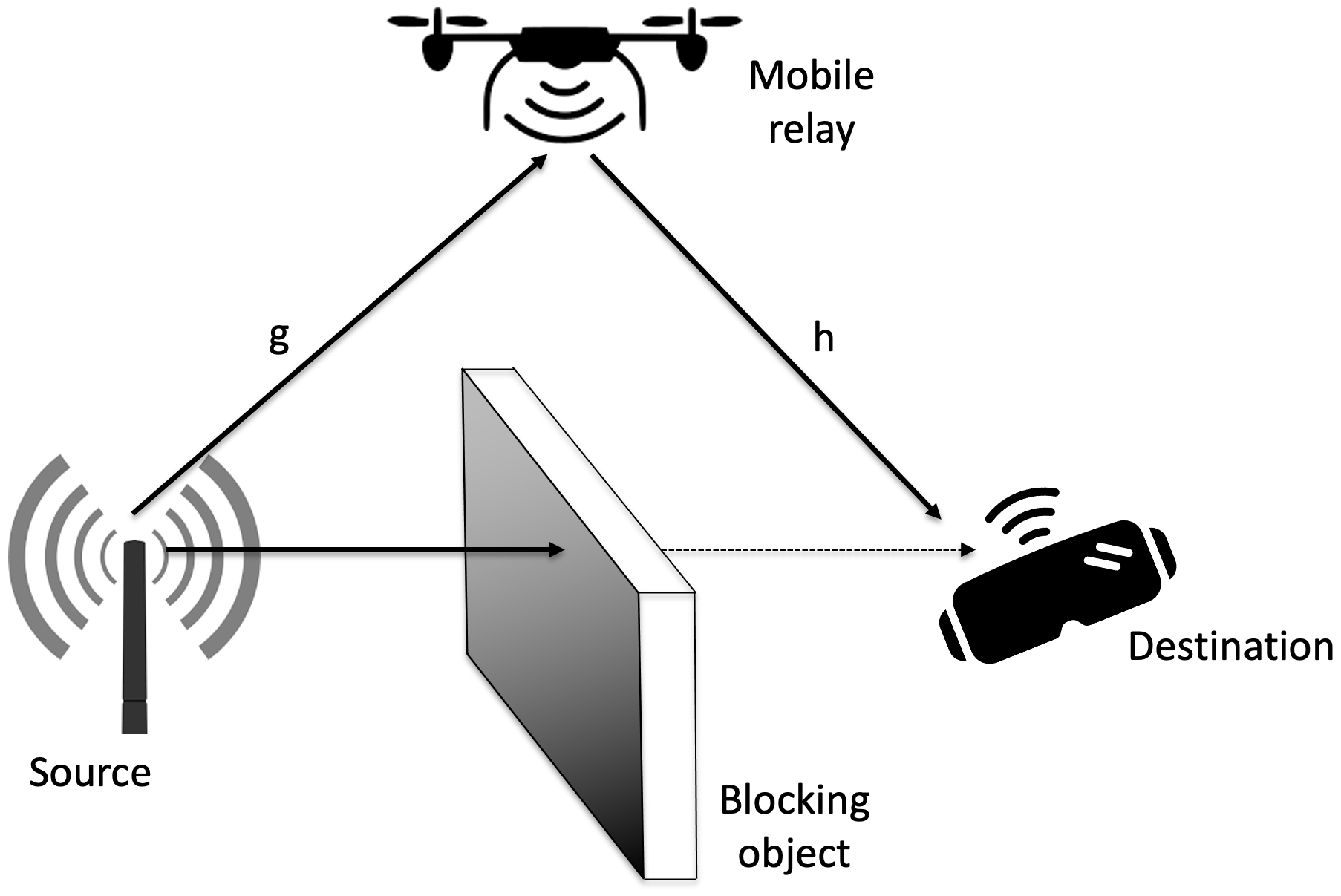}
\vspace{-1mm}
\caption{MR-supported NLoS avoidance}
\label{fig:setup_mr}
\vspace{-3mm}
\end{figure}  

The transmission with this type of \ac{MR} is carried out in two phases, first from the source to the \ac{MR}, followed by the retransmission from the \ac{MR} to the destination, assuming no \ac{LoS} transmission~\cite{levin2012amplify}.
Therefore, the \ac{SNR} from the source to the \ac{MR}, as well as the \ac{SNR} from the \ac{MR} to the destination, can be expressed using respectively Equations~\eqref{eq1} and~\eqref{eq2}, while the \ac{SNR} that corresponding to the end-to-end communication link can be expressed as $\min{(SNR_{(Tx,MR)},SNR_{(MR,Rx)})}$.

\begin{equation}
\label{eq1}
SNR_{(Tx,MR)} = \frac{P_{Tx}}{|\sigma|^2}\sum_{n=1}^N{|h_n|^2},
\end{equation}

\begin{equation}
\label{eq2}
SNR_{(MR,Rx)} = \frac{P_{MR}}{|\sigma|^2}\sum_{n=1}^N{|g_n|^2}.
\end{equation}

In the above, $h_n$ represents the channel from the source to the \ac{MR}, $g_n$ represents the channel from the \ac{MR} antenna to the destination, while $n \sim N_c(0,\sigma^2)$ is the noise at the destination~\cite{bjornson2020power}. 
In this case, the \ac{SNR} grows with the number of antenna elements $N$~\cite{bjornson2020power,lee2002large}.
The channel gain for $g_n$ and $h_n$ can be expressed by:

\footnotesize
\begin{dmath}
\zeta_{(d,\eta,N)} = \sum_{i=1}^{2}\left(\frac{B+(-1)^i\sqrt{B}\tan{\eta}}{6\pi(B+1)\sqrt{2B+\tan^2{\eta}+1+2(-1)^i\sqrt{B}\tan{\eta}}}+\frac{1}{3\pi}\tan^{-1}\frac{B+(-1)^i\sqrt{B}\tan{\eta}}{\sqrt{2B+\tan^2{\eta}+1+2(-1)^i\sqrt{B}\tan{\eta}}}\right),
\end{dmath}
\normalsize

where $B$ is equal to $N A/(4d^2 \cos^2{\eta})$, $d$ is the distance between two directly communicating devices, while $\eta$ is the angle between the source and the destination from the MR perspective.
To calculate the position of the source and the destination with respect to the \ac{MR}, we follow the same principle as before for the \ac{IRS}-supported \ac{NLoS} avoidance.

\section{System Analysis}
\label{sec:analysis}

Our motivating scenario comes from the domain of full-immersive multi-user VR systems such as the one presented in~\cite{struye2021millimeter}, in which directional mmWave beams are expected to “track” the users’ movements for maintaining \ac{LoS} connectivity, hence maximizing the quality of video delivery.
However, the users could block their own \ac{LoS} links with their bodies or they could be interrupted by the other users.
In such cases, we envision the \ac{NLoS} avoidance to be supported through the utilization of either the IRS or MR. 

In the analysis of the IRS- and MR-supported NLoS avoidance, we assume the environment is defined as a square of a certain size. 
In the first scenario, we assume the \ac{IRS} can be placed on any of the outer walls in the environment, so in our scenario the IRS could be positioned anywhere on the sides of the square, while the floor and the ceiling have not been considered for simplicity. 
In the second scenario, the \ac{MR} will be utilized for NLoS avoidance, where we assume the \ac{MR} can be placed anywhere in the environment.
We assume both the IRS and the MR to be placed in a 2D environment, however also accounting for the height of the IRS in the derivation of the communication link quality indicators. 

\begin{figure}[!t]
\centering
\includegraphics[width=\columnwidth]{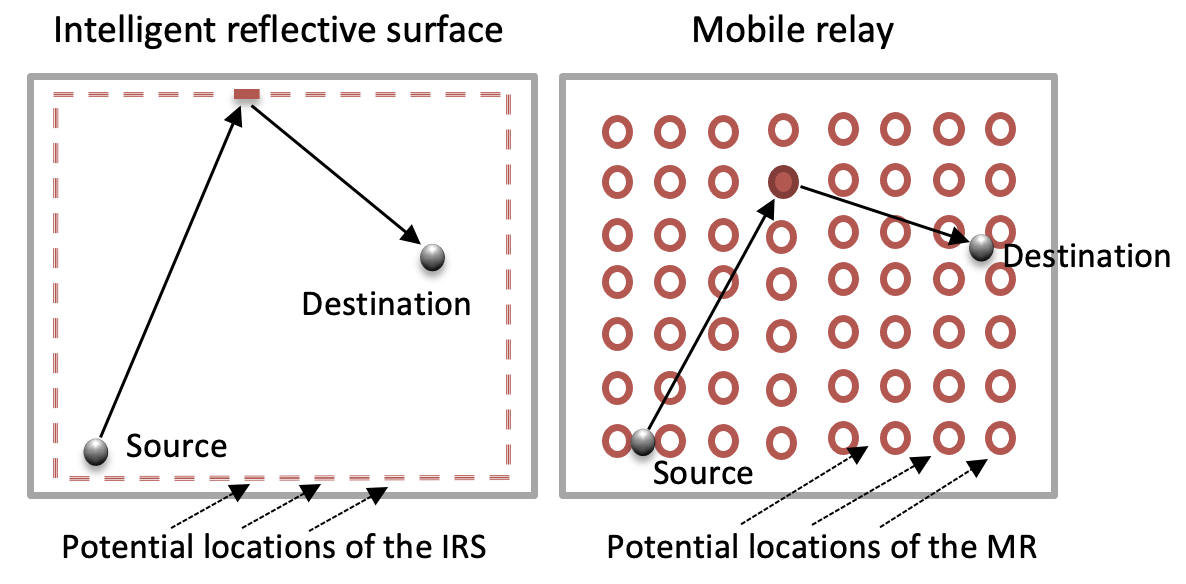}
\vspace{-4mm}
\caption{Approach for deriving the optimal positions of the IRS and MR in the considered scenario}
\label{fig:algorithm}
\vspace{-5mm}
\end{figure}

These two scenarios are depicted in Figure~\ref{fig:algorithm}, jointly with the indication of the potential positions of the IRS and MR. 
To calculate the optimal position of the \ac{IRS} for a particular position of the source and the destination, the \ac{SNR} is calculated for all of the potential positions of the \ac{IRS} on the four walls of the environment with respect to the source and the destination. 
To specify the set of positions the \ac{IRS} could take, we utilize the approach of exhaustively iterating along the four walls of the environment with an arbitrary step distance. 
To calculate the optimal position of the \ac{MR} for a specific position of the source and the destination, the \ac{SNR} is calculated as in the previous scenario for a grid-like constellation of possible \ac{MR} positions, where the distance between two neighboring potential positions features the same arbitrary step distance.

The \ac{SNR} of the communication link between the source and the destination at each location of the IRS or MR can be calculated using Equations~\ref{eq:snr_irs},~\ref{eq1}, and~\ref{eq2}, which take their ground from~\cite{bjornson2020power,bjornson2021reconfigurable}.
The derivation of the optimal locations of the IRS and MR is based on the calculation of the \ac{SNR} for all potential positions of the IRS and MR.
For each source and the destination pair, the position with the highest \ac{SNR} is considered as the optimal position of the \ac{IRS} and \ac{MR}. 
The only difference in the \ac{IRS}- and \ac{MR}-focused scenarios, apart from the approach for calculating the \ac{SNR}, is that for the \ac{MR} the entire environment is to be considered, while for the \ac{IRS} positioning only the four outer walls of the environment are considered as feasible.


\vspace{-1mm}
\section{Evaluation Results}
\label{results}

\subsection{Evaluation Setup}

In this section, we outline the simulation approach and setup used for establishing and comparing the link quality of IRS- and MR-supported NLoS avoidance in indoor mmWave networks. 
Given that we envision the utilization of mmWave frequencies in the range between 30 and 100~GHz~\cite{nibir2020quintuple,lu2019integrated}, we have selected 30~GHz as the operating frequency. 
As such, the communication wavelength $\lambda$ was set to 0.01~m.
Based on~\cite{wu201764}, the noise figure was set to 6~dB, antenna gain at the source to 6~dBi, and communication bandwidth to 1.76~GHz. 
Moreover, we have considered an additive white Gaussian noise affecting the communication performance of both \ac{IRS} and \ac{MR}-based setups, whose power equals to $N = -174+10 \log(B)+F_{dBm}$, where \emph{B} is the communication bandwidth~\cite{ntontin2019multi} and $F_{dBm}$ is the noise figure. 

The simulation environment was assumed to be a square with sizes of $3\times3$, $6\times6$ and $10\times10$~m$^2$, with the objective of simulating three spaces that correspond to real scenarios. 
Specifically, we aimed at mimicking an apartment room corresponding to the size of $3\times3$~m$^2$, while an office and a conference room would correspond to $6\times6$ and $10\times10$~m$^2$.

The area of each antenna element was set to $A = (\lambda/4)^2$, where the side length of each element equals $a = \sqrt{A}$~\cite{bjornson2020power}.
Moreover, we have considered three sets of antenna elements for both the IRS- and MR-based scenarios, with the numbers of antenna elements equaling to 100, 400, and 900. 
The reason for considering different numbers of antenna elements was to assess the benefits of increasing antenna array sizes on the performance in the \ac{IRS}- and \ac{MR}-supported NLoS avoidance.
The above numbers of antenna elements for \ac{IRS}-supported communication have been selected according to the existing literature, e.g.,~\cite{bjornson2020reconfigurable}.
The same numbers have been utilized for \ac{MR}-supported communication for supporting objective comparison between the two scenarios, ignoring the practicality of utilizing large numbers of antennas by the MR.

The complete analysis has been carried out over 10,000 simulation runs.
In each run, the locations of the source and destination have been randomly defined in the environment.  
This was followed by deriving the optimal locations of the IRS and MR by utilizing the iterative approach presented in Section~\ref{sec:analysis}.  
Specifically, for the IRS-based scenario the optimal location search space consisted of the four outer walls of the specified environments, with the granularity of the neighboring potential locations (i.e., step distance) equaling 10~cm.
For the MR-based scenario, the location search space was constrained to the interiors of the specified environments, with the granularity of the neighboring locations again equaling 10~cm.

\begin{figure*}[!t]
\centering
\includegraphics[width=.78\linewidth]{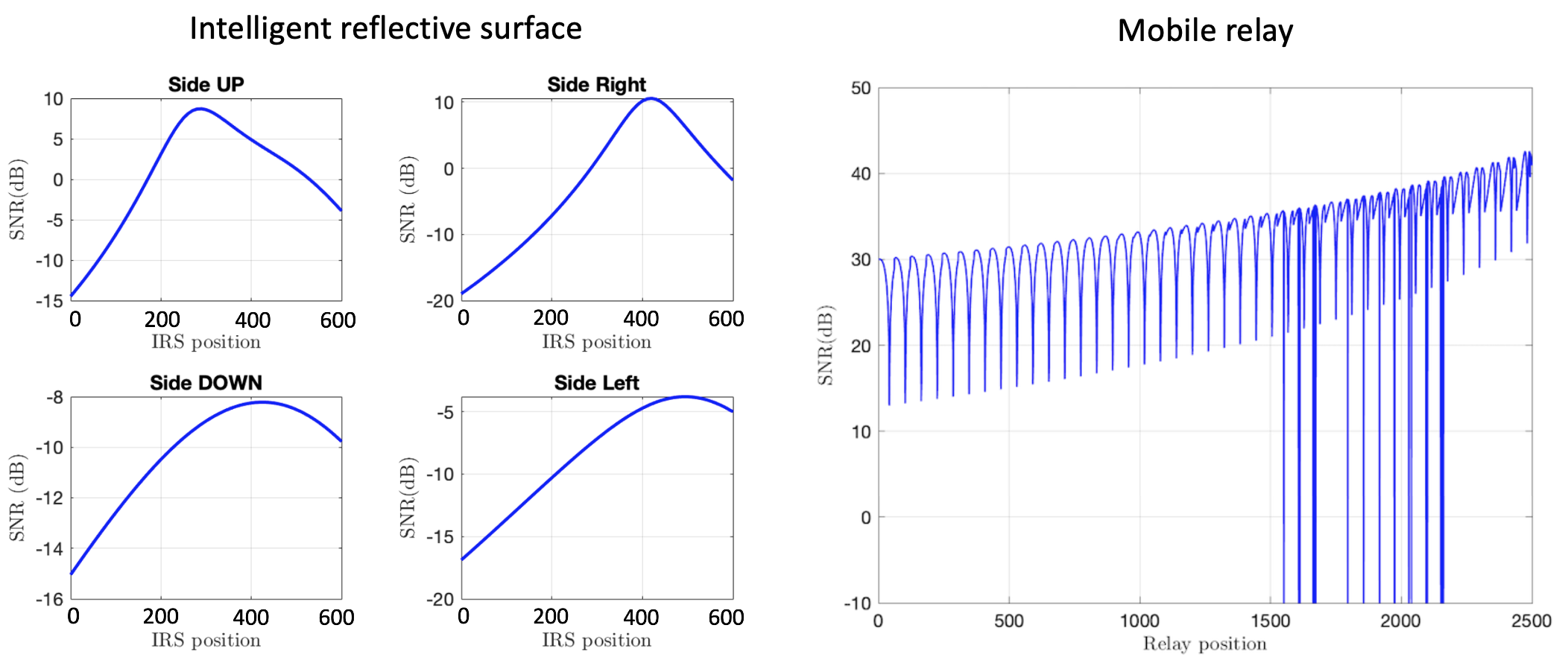}
\vspace{-2mm}
\caption{Example derivation of the optimal IRS/MR location}
\label{fig:iteration}
\vspace{-5mm}
\end{figure*}

An example derivation of the optimal positions of the IRS and MR for a single position of the source and the destination is given in  Figure~\ref{fig:iteration}.
As visible in the figure, for the IRS-based scenario where the optimal location search space is constrained to the outer walls of the simulation environment, the optimal location of the IRS (i.e., the one that maximizes the SNR of the communication link between the source and the destination) is found to be on the right wall of the environment.
Similarly, for the MR-based scenario the optimal location search space is specified in a grid-like fashion across the entire simulation environment, while the optimal location is the one that yields the highest SNR of the communication link.
This is with the exception of the potential relay locations at which the SNR of the communication link was higher than the SNR of the direct LoS communication between the source and destination.
In such cases, the SNR of the communication link through the MR is considered to be infeasible, hence these locations have been discarded by setting the corresponding SNR to -10~dBm, as visible in the figure.		

\subsection{Evaluation Results}

\begin{figure*}[!t]
\centering
\includegraphics[width=0.8\linewidth]{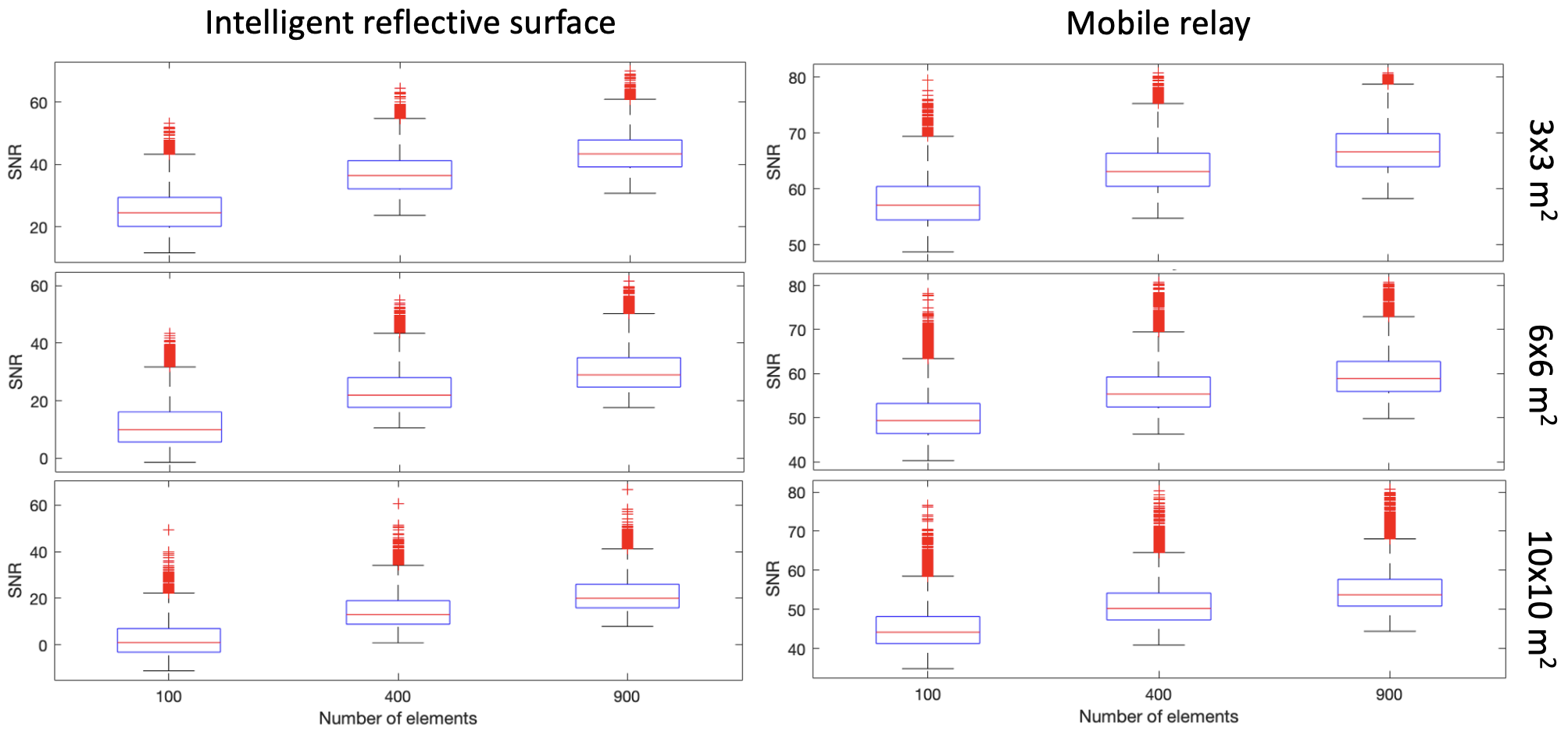}
\vspace{-1mm}
\caption{Average SNRs for IRS- and MR-supported communication}
\label{fig:comparison}
\vspace{-5mm}
\end{figure*}

The communication quality for the IRS- and MR-supported NLoS avoidance in the considered scenarios, characterized using the average SNR over 10,000 simulation runs, is depicted in Figure~\ref{fig:comparison}.  
The obtained SNRs are depicted in the standard box-plot fashion for different environment sizes and numbers of antenna elements, as indicated in the figure.
For the environment sizes of $3\times3$~m$^2$ and assuming the \ac{IRS}-supported communication, it can be observes that the maximum \ac{SNR} increases by approximately 40 and 80\% when the number of IRS antenna elements increases from 100 to 400 and from 100 to 900, respectively.
Assuming the \ac{MR}-supported communication, the figure shows that the average \ac{SNR} increases approximately 10\% when the number of antennas increases from 100 to 400, and approximately 18\% when the number of antennas increases from 100 to 900.

Similar observations can be made for different environment sizes. 
Specifically, if we analyze the environment with sizes of $6\times6$~m$^2$ and assuming the \ac{IRS}-supported communication, there is an approximate 50\% increase in the average SNR quality when increasing the number of antenna elements from 100 to 400, as well as an additional 30\% increase when the number of antenna elements is increased from 400 to 900. 
For \ac{MR}-supported communication, the average \ac{SNR} increases by respectively 10 and 20\% if the number of antenna elements is increased from 100 to 400 and 900.
For the environment with sizes of $10\times10$~m$^2$ and the \ac{IRS}-supported communication, an approximate increase of 8~dB is observed when increasing the number of antenna elements from 100 to 400 and an additional one of 10~dB if the increase is from 400 to 900 elements. 
When the \ac{MR} is used with 400 antenna elements, an increase is approximately 5~dB, while with 900 antennas approximately 10~dB compared to the antenna array of 100 elements. 

When comparing the communication link performance as a function of the sizes of the considered environments, it can be observed that the increase in the sizes of the environment degrades the average quality of the communication links.
For example, when utilizing the \ac{IRS} with 100 antenna elements the average \ac{SNR} is reduced with respect to the $3\times3$~m$^2$-sized environment by 32\% and 55\% as the environment sizes increase to respectively $6\times6$~m$^2$ and $10\times10$~m$^2$. 
Similarly, the average \ac{SNR} is reduced with respect to the $3\times3$~m$^2$-sized environment by 34\% and 12\% with the use of the \ac{MR}.
Similar trends can be observed for different configurations pertaining to the number of utilized antenna elements in both IRS and MR-supported communication, as visible in Figure~\ref{fig:comparison}.

Based on the derived results, we can observe that the \ac{MR} generally yields significantly better communication performance than the \ac{IRS}, with the average improvement of roughly 20~dB observed along all parameterizations of the scenario of interest.
We argue the reason can be found in the fact that the MR actively retransmits the signals, while the IRS passively reflects them with phase-shifts. 
Moreover, the MR features a higher degree of freedom in terms of its optimal positioning in the environment of interest.
Due to that, the average distance that signals have to traverse between the source and destination are lower for MR-supported communication compared to its IRS-supported counterparts. 

Besides that, our results demonstrate that, if the design goal is to guarantee a certain link quality of an indoor mmWave communication system in NLoS conditions using the IRS or MR, the sizes of the environment should be taken into account jointly with the number of antenna elements deployed at the IRS or MR level.
In other words, increasing the number of antenna elements will be needed for coping with the signal attenuation at mmWave frequencies due to the increased communication ranges stemming from the increase in the sizes of the environment. 
Finally, our results also motivate the need for optimal positioning of the \ac{IRS} or MR for optimizing the communication quality between the source and the destination.
The motivation for that can be found in Figure~\ref{fig:iteration} demonstrating that the quasi-optimal positioning (i.e., optimal for the given granularity between potential locations of the IRS and MR) of both the \ac{IRS} and \ac{MR} can increase the SNR of communication by more than 25~dB compared with the worst case scenario. 

\section{Conclusion}

We have considered the \acfp{IRS} and \acfp{MR} as potential technologies for mitigating the negative effects of \acf{NLoS} connectivity in indoor mmWave networks. 
Accounting for the characteristics of this specific setup, the MRs achieve significantly higher average \acf{SNR} of communication than the IRSs. 
This is a direct consequence of the higher degree of freedom of movement that the MRs feature compared to the IRSs. 
Nonetheless, one should realize that the IRSs have intrinsic advantages that could make them preferable to the MRs, such as their lower cost and higher energy efficiency. 
Therefore, our indications currently only pertain to the communication quality indications and are tightly related to the freedom of positioning of the IRSs and MRs.
Future work will be focused on deriving a comparison framework that is able to capture different performance metrics including production and maintenance costs, energy consumption, communication reliability and latency. 
We will also consider utilizing a faster optimization approach for IRS and MR positioning than the exhaustive search, as well as 3D environments and scenarios with multiple users. 
Finally, in this work we have assumed that the MRs can instantaneously reach their optimal positions, which will in reality incur a certain delay, effectively reducing the throughput of communication.
Therefore, part of the future work will be focused on incorporating such delays in the MR-focused scenario and capturing their effects on the performance of MR-supported NLoS avoidance in mmWave communication.

\vspace{-0.5mm}
\section*{Acknowledgments}

Filip Lemic was supported by the EU H2020 Marie Skłodowska-Curie project ”Scalable Localization-enabled In-body Terahertz Nanonetwork” (nr. 893760).

\renewcommand{\bibfont}{\footnotesize}
\printbibliography

\end{document}